\documentstyle[epsfig]{ichep}

\def\lapproxeq{\lower .7ex\hbox{$\;\stackrel{\textstyle <}{\sim}\;$}}
\def\gapproxeq{\lower .7ex\hbox{$\;\stackrel{\textstyle >}{\sim}\;$}}

\begin{document}

\title{\hfill DTP/94/76\break Low $x$ phenomena}
\author{A.D.\ Martin}

\address{Department of Physics, University of Durham, Durham DH1 3LE,
England.}

\abstract{We review recent developments in the application of perturbative QCD
to phenomena at small $x$.}

\twocolumn[\maketitle]
\fnm{7}{To be published in the Proc. of ICHEP Conf., Glasgow, July 1994.}

Both H1 \cite{H1} and ZEUS \cite{ZEUS} presented measurements of $F_2(x,Q^2)$
obtained from the 1993 HERA run.  A sample of these data is shown in Fig.\ 1.
The dramatic rise of $F_2(x,Q^2)$ with decreasing $x$, discovered in the 1992
data, is now firmly established.  The data are compatible with Altarelli-Parisi
(or GLAP) evolution from \lq\lq starting" parton distributions in which the sea
quarks have the small $x$ behaviour
\begin{equation}
x{\cal S} \sim x^{-\lambda} \;\; {\rm with} \; \lambda \sim 0.3,
\end{equation}
see, for example, the MRS(A) curve \cite{MRSA} in Fig.\ 1.

Also OPAL \cite{OPAL} presented a measurement of the photon structure function
$F_2$ at $x$ = few $\times 10^{-2}$.  No rise with decreasing $x$ is seen at
this value of $x$.  However both HERA and LEP2 have the potential to measure
the
photon structure function at much smaller $x$.  Here we concentrate on the
structure of the proton since many results  have already been obtained in the
HERA small $x$ regime $(x \lapproxeq 10^{-3})$.  \section{Deep-inelastic map}

To orientate ourselves we show in Fig.\ 2 a map of the kinematic regime for
deep
inelastic electron-proton scattering.

\begin{figure}[b]
\begin{center}
\end{center}
\caption{A sample of the latest HERA data for $F_2$ [1,2].  The GLAP-based
MRS(A) analysis [3] and the BFKL-based AKMS \lq\lq prediction" [4] give almost
indistinguishable descriptions of the HERA data.  Also shown are the GRV parton
[5] and \lq\lq hot-spot" shadowing [6] predictions.}
\label{1}
\end{figure}

\vspace*{.5cm}

\noindent {\bf (a) GLAP evolution (large $Q^2$)}

Starting from a known structure of the proton at $Q^2 = Q^2_0$ we may evolve up
to large log$Q^2$ using the Altarelli-Parisi (or GLAP) equations which are
typically of the form
\begin{equation}
\partial g/\partial {\rm log}Q^2 = P_{gg} \otimes g + ...
\end{equation}
where the convolution is over the longitudinal momentum fraction.  For
simplicity, we concentrate on the gluon, the dominant parton at small $x$.
Effectively the GLAP equations resum the leading $(\alpha_s{\rm log}Q^2)^n$
contributions, which correspond (in a physical gauge) to the $n$-rung gluon
ladder of Fig.\ 3.  In fact the leading log arises from the strongly-ordered
region of transverse momenta
\begin{equation}
Q^2 \gg k^2_{Tn} \gg k^2_{Tn-1} \gg ...
\end{equation}
When we evolve to high $Q^2$ we probe the proton structure ever more finely, to
transverse sizes $\sim 1/\sqrt{Q^2}$, see Fig.\ 2.

\begin{figure}[t]
\begin{center}
\end{center}
\caption{The gluonic content of the proton as \lq\lq seen" in different
deep-inelastic $(x,Q^2)$ regions.  $W$ is the ratio of the quadratic to the
linear term on the right hand side of (10).}
\label{2}
\end{figure}

The parton distributions are essential to calculate the cross sections,
$\sigma$, for \lq\lq hard" hadronic processes.  First the QCD subprocess,
$\hat{\sigma}$, are calculated in the strongly-ordered $k^2_T = 0$
approximation
and then the factorization theorem gives  $\sigma = xg(x,\mu^2) \otimes
\hat{\sigma} (\mu^2) + ...$ where the scale $\mu^2 \sim$ hard scattering
$p^2_T$.

\vspace*{.25cm}

\noindent {\bf (b) BFKL equation (small $x$)}

On the other hand, when we evolve up to large $1/x$ (i.e.\ small $x$) we
encounter ($\alpha_s$ log$(1/x))^n$ terms which have to be resummed.  Indeed
the
dramatic rise observed in $F_2$ with decreasing $x$ may be associated with the
growth of the gluon density which arises from the resummation of these terms; a
growth which, via $g \rightarrow q\bar{q}$, is transmitted to the sea quarks
probed by the photon.  To leading order the summation is accomplished by the
BFKL (or Lipatov) equation \cite{BFKL}, which may be written in the
differential
form
\begin{displaymath}
-x \partial f(x,k^2_T)/\partial x =
\end{displaymath}

\begin{displaymath}
\frac{3\alpha_s}{\pi} k^2_T \int^{\infty}_0 \frac{dk^{\prime 2}_T}{k^{\prime
2}_T}\left[ \frac{f(x,k^{\prime 2}_T) - f(x,k^2_T)}{|k^{\prime 2}_T - k^2_T|} +
\frac{f(x,k^2_T)}{(4k^{\prime 4}_T - k^4_T)^{\frac{1}{2}}} \right]
\end{displaymath}

\begin{equation}
\equiv K \otimes f .
\end{equation}
There is now no strong-ordering in $k_{nT}$ of the emitted gluons and we have
to
work in terms of the unintegrated gluon distribution $f(x,k^2_T)$ in which the
\lq\lq last" $k^2_T$ integration along the gluon ladder (of Fig.\ 3a) is
unfolded, that is

\begin{figure}[b]
\begin{center}
\end{center}
\caption{(a) Gluon ladder, (b) diagrammatic representation of the BFKL
contribution
to $F_2$, see (13).}
\label{3}
\end{figure}

\begin{equation}
xg(x,\mu^2) = \int^{\mu^2} \frac{dk^2_T}{k^2_T} f(x,k^2_T) .
\end{equation}
At small $x$ the gluon distribution $f(x,k^2_T,\mu^2)$ becomes independent of
the scale $\mu^2$.

 From (4) we see that the small $x$ behaviour of $f$ is controlled by the
largest
eigenvalue $\lambda_L$ of the eigenfunction equation $K \otimes f_n =
\lambda_nf_n$, since as $x \rightarrow 0$
\begin{displaymath}
f \sim {\rm exp}(\lambda_L {\rm log}(1/x)) \sim x^{-\lambda_L} .
\end{displaymath}
Indeed for fixed $\alpha_s$ there is an analytic solution for the leading small
$x$ behaviour
\begin{equation}
f \sim x^{-\lambda_L} (k^2_T)^{\frac{1}{2}} {\rm exp} \left( -c \frac{{\rm
log}^2(k^2_T/\bar{k}^2_T)}{{\rm log}(1/x)} \right)
\end{equation}
where
\begin{equation}
\lambda_L = (3\alpha_s/\pi)4 {\rm log} 2 .
\end{equation}
This singular $x^{-\lambda_L}$ Lipatov behaviour is in contrast to the naive
Regge-type expectations that
\begin{equation}
f \sim x^{1-\alpha_P(0)} \sim x^{-0.08}
\end{equation}
where $\alpha_P(0)$ is the intercept of the Pomeron.

A second feature of the solution (6) of the BFKL equation is the diffusion in
$k_T$ with decreasing $x$, as manifested by the Gaussian form in log$k^2_T$
with
a width which grows as (log$(1/x))^{\frac{1}{2}}$ as $x$ decreases.  The
physical origin of the diffusion is clear.  Since there is no strong-ordering
in
$k_T$, there is a \lq\lq random walk" in $k_T$ as we proceed along the gluon
chain and hence evolution to smaller $x$ is accompanied by diffusion in $k_T$.
We foresee that the diffusion will be a problem in the applicability of the
BFKL
equation since, with decreasing $x$, it leads to an increasingly important
contribution from the infrared and ultraviolet regions of $k^2_T$ where the
equation is not expected to be valid.

For running $\alpha_s$ the singular behaviour and diffusion in $k_T$ are
confirmed.  In addition it is found that \cite{AKMS}
\begin{equation}
f \sim C(k^2_T) x^{-\lambda}
\end{equation}
where the value $\lambda \approx 0.5$ is much less sensitive to the treatment
of
the infrared region in (4) than is the normalization $C$.

\vspace*{.25cm}

\noindent  {\bf (c) Shadowing region}

The $x^{-\lambda}$ growth of the gluon cannot go on indefinitely with
decreasing
$x$.  It would violate unitarity.  The growth must eventually be suppressed by
gluon recombination, which is represented by an additional quadratic term so
that (4) has the symbolic form
\begin{equation}
-x \partial f/\partial x = K \otimes f - V \otimes f^2 .
\end{equation}
The additional term contains a factor $\alpha^2_s/k^2_T R^2$ since the
gluon-gluon
interaction behaves $\sim \alpha^2_s/k^2_T$, whereas $1/R^2$ arises because the
smaller the transverse area $(\pi R^2)$, in which the gluons are concentrated
within the proton, the stronger the effect of recombination.  The precise form
of this equation, originally proposed by GLR \cite{GLR}, is still a matter of
debate \cite{BAR}.  The region where shadowing should be calculable
perturbatively is just below the dashed line in Fig.\ 2.
\section{Small $x$ behaviour of $F_2$}

The small $x$ behaviour of $xg$ (and $x\bar{q}$) arising from GLAP evolution
depends on the form of the starting distributions.  For singular starting
distributions, $xg \sim x^{-\lambda}$ with $\lambda > 0$, the small $x$
behaviour is stable to evolution in $Q^2$.  The larger the value of $\lambda$
the sooner the stability sets in with decreasing $x$.  MRS(A) partons
\cite{MRSA}, with $xg \sim x^{-0.3}$, are an example of this behaviour.  On the
other hand, for non-singular starting distributions, $xg \sim x^{-\lambda}$
with
$\lambda \leq 0$, we find   the double leading logarithm (DLL) form
\begin{equation}
xg \sim {\rm exp} (2[\xi (Q^2_0,Q^2){\rm log} (1/x)]^{\frac{1}{2}}).
\end{equation}
That is $xg$ grows as $x \rightarrow 0$ faster than any power of log$(1/x)$ but
slower than any power of $x$.  The larger the \lq\lq evolution length",
\begin{equation}
\xi = \int^{Q^2}_{Q^2_0} \frac{dq^2}{q^2} \frac{3\alpha_s(q^2)}{\pi} ,
\end{equation}
the faster the growth.  An example is the \lq\lq dynamical" GRV partons
\cite{GRV} which evolve from valence-like forms at a low scale $Q^2_0 = 0.3$
GeV$^2$, and for which (11) mimics a behaviour $xg \sim x^{-0.4}$ in the HERA
regime.

Given the solution $f(x,k^2_T)$ of the BFKL equation we can use the
$k_T$-factorization theorem to predict $F_2$, see Fig.\ 3b:
\begin{equation}
F_2 = f \otimes F^{{\rm box}} + F^{{\rm bg}}_2 \simeq C^{\prime}(k^2_T)
x^{-\lambda} + F^{{\rm bg}}_2
\end{equation}
where $\lambda \simeq 0.5$, and $F_2^{{\rm bg}} \simeq 0.4$ is determined from
the large $x$ behaviour of $F_2$.  Once the overall normalisation of the BFKL
term is adjusted by a suitable choice of the infrared parameters in (4), then
an
excellent description of all the $F_2(x,Q^2)$ HERA data is obtained.  Indeed
the
BFKL-based \lq\lq prediction" \cite{AKMS} gives an equally good, and almost
indistinguishable, description as the GLAP-based order fit
\cite{MRSA}, see Fig.\ 1.  With GLAP, the steepness is either incorporated (as
a
factor $x^{-\lambda}$) in the starting distributions or generated by evolution
from a low scale $Q^2_0$.  The steepness can be adjusted to agree with the data
by varying $\lambda$ or $Q^2_0$.  On the other hand the leading log$(1/x)$ BFKL
prediction for the shape $F_2-F^{{\rm bg}}_2 \sim x^{-\lambda}$, with $\lambda
\simeq 0.5$, is prescribed.  It remains to see how well it survives a full
treatment of sub-leading effects.

Conventional shadowing with gluons spread uniformly across the proton $(R = 5$
GeV$^{-1}$) leads to only a small suppression in $F_2$ in the HERA regime.  If
the gluons were concentrated in \lq\lq hot spots" of area $\pi R^2$ with, say,
$R = 2$ GeV$^{-1}$ the effect would be much stronger, see Fig.\ 1.  But could
shadowing be identified since we do not know the partons at small $x$?
Simulated $F_2$ data (of accuracy and $x$ range which may eventually be
accessible at HERA) have been used \cite{GKR} to see how well $R$ could be
determined.  The conclusion is that the interplay between the linear and
non-linear terms in (10) leads to a considerable ambiguity between the size of
the parton distributions  and the amount of shadowing.

Is GLAP evolution adequate in the HERA regime?  For sufficiently small $x$ the
$(\alpha_s$log$1/x)^n$ terms must be resummed with the full $Q^2$ dependence
(and not just the leading and next-to-leading log$Q^2$ terms).  Ellis et al.\
\cite{EKL} have made a theoretical study of the applicability of GLAP evolution
and find that it is adequate in the HERA small $x$ regime, {\it provided} that
the evolution occurs from a sufficiently singular starting distribution, $xg
\sim x^{-\lambda}$ with $\lambda \gapproxeq 0.35$.

\section{Identification of BFKL behaviour}

The {\it inclusive} nature of $F_2$, and the necessity to provide \lq\lq
non-perturbative" input distributions of parton densities for its description,
prevents its observed small $x$ behaviour being a sensitive discriminator
between BFKL and conventional dynamics.  For this purpose it is necessary to
look into the properties of the final state.

The two characteristic features of BFKL dynamics are the absence of
strong-ordering of the gluon $k_T$'s along the chain (the
diffusion in $k_T$) and the consequent $(x/x^{\prime})^{-\lambda}$ or
exp$(\lambda\Delta y)$ growth of the cross section, where $x$ and $x^{\prime}$
are the longitudinal momentum fractions of the gluons at the ends of the chain,
which spans the rapidity interval $\Delta y = {\rm log}(x^{\prime}/x)$.  Recall
$\lambda \simeq 0.5$.  Some processes which exploit these characteristic
features are shown in Fig.\ 4.

\begin{figure}[t]
\begin{center}
\end{center}
\caption{Processes that may be used to identify BFKL dynamics.}
\label{4}
\end{figure}

The idea \cite{M} in Fig.\ 4a is to detect deep-inelastic $(x,Q^2)$ events
which
contain a measured jet $(x_j,k^2_{Tj})$ in the kinematic regime where (i) the
jet longitudinal momentum, $x_j$, is as large as is experimentally feasible
$(x_j \sim 0.1)$, (ii) $z \equiv x/x_j$ is small, and (iii) $k^2_{Tj} \approx
Q^2$ is
sufficient to suppress diffusion into the infrared region.  The beauty of this
measurement is that attention is focussed directly on the BFKL $z^{-\lambda}$
behaviour at small $z$.  The difficulty is to cleanly separate the forward
going
jet from the proton remnants.  The preliminary results from the H1
collaboration
\cite{H12} are encouraging and favour the BFKL over the conventional
description.

Inspection of Fig.\ 4b suggests, that due to the relaxation of strong-ordering
of the gluon $k_T$'s at small $x$, more transverse energy $E_T$ should be
emitted in the central region (between the current jet and the proton remnants)
than would result from conventional evolution.  Indeed Monte Carlo predictions
based on QCD cascade models\footnote{A good description has been obtained by a
Monte Carlo based on the colour dipole model.  This model contains the essence
of BFKL dynamics
provided the full integration over the emitted gluon $k_T$ is performed.} fall
well below the observed central plateau of height $E_T \approx 2.1$ GeV per
unit
of rapidity \cite{H13}.  A BFKL-based calculation \cite{GKMS}, at the parton
level, yields about 1.7 GeV per unit of rapidity, but much less if conventional
dynamics is used.  No hadronization effects have been allowed for.

Recently there has been renewed interest in the original proposal of Mueller
and
Navelet \cite{MN} that the cross section for the production of a pair of
minijets should, according to BFKL dynamics, rise as exp$(\lambda \Delta y)$ as
the rapidity interval $\Delta y$ becomes large.  The studies \cite{DSS} show
that the effect is masked by the fall-off of the parton densities at large $x$,
but that instead the rate of weakening of the azimuthal (back-to-back)
correlation between the jets, could possibly be an indicator of BFKL effects.

BFKL dynamics may be also identified via the weakening of the azimuthal
correlation between a pair of jets produced in deep-inelastic scattering at
HERA, see Fig.\ 4d.  At sufficiently large values of $\Delta \phi \equiv \phi -
\pi$, BFKL dynamics dominates over the fixed-order QCD contribution from 3+1
jet
production, leading to a distinctive tail in the azimuthal distribution which
directly probes the $k_T$ dependence of the gluon distribution \cite{AGKM}.

\section{Conclusions}

In the HERA small $x$ regime GLAP evolution (from appropriately parametrized
starting distributions) is able to mimic BFKL dynamics as far as the
description
of $F_2(x,Q^2)$ is concerned.  Moreover it will be difficult to isolate
shadowing contributions even with improved measurements of $F_2$.  Measurements
which are less inclusive than $F_2$, offer more chance to identify the
characteristic BFKL $x^{-\lambda}$ behaviour and diffusion in $k_T$.  However
opening up the final state brings the problems of hadronization and jet
identification, and loses some of the small $x$ \lq\lq reach" (e.g. $x
\rightarrow x/x_j$ where $x_j \sim 0.1$).  On the theoretical side, the
sub-leading corrections to the BFKL leading log$(1/x)$ formalism are urgently
needed for future quantitative studies of small $x$ phenomena.

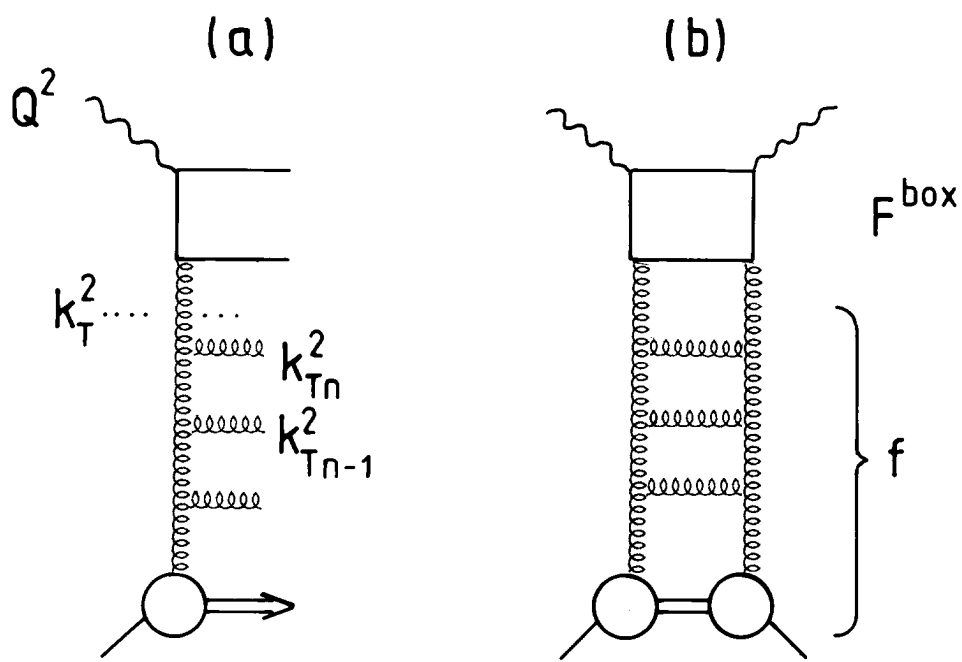
\end{document}